\documentstyle[12pt,graphicx]{article}
\begin{document}
\title{Slow light in moving media}
\author{U.\ Leonhardt\\
School of Physics and Astronomy\\ University of St Andrews\\
North Haugh\\ St Andrews, Fife, KY16 9SS\\ Scotland \and P.\
Piwnicki \\ Physics Department\\ Royal Institute of Technology
(KTH)\\ Lindstedtsv\"agen 24\\ S-10044 Stockholm\\ Sweden}
\maketitle
\begin{abstract}
We review the theory of light propagation in moving media with
extremely low group velocity. We intend to clarify the most
elementary features of monochromatic slow light in a moving medium
and, whenever possible, to give an instructive simplified picture.
\end{abstract}

\newpage
\section{Introduction}

Waves experience moving media as effective gravitational fields.
To understand why, imagine a definite example --- light traveling
in a moving transparent fluid such as flowing water
\cite{LPstor,LPcp}. In each drop of the liquid, light propagates
along a straight line and so each drop distinguishes a particular
inertial frame. Now imagine light passing from one drop to a next
one which happens to move with a different velocity vector.
Again, the new drop in the way of the light distinguishes an
inertial frame, but this new frame will differ from the previous
one. Light traveling in a non--uniformly moving medium is
constantly forced to adapt to new inertial frames. An analogous
situation occurs in curved space--time when light, and all other
matter, experiences space and time as consisting of connected
inertial frames. Space--time can be thought as being locally
flat, even in the vicinity of the most violent gravitating
objects, yet each local frame is non-trivially connected to the
neighboring frames, a connection mathematically described in
terms of the curvature tensor \cite{LL2}. In a moving medium, the
local co-moving frames of the medium play the role of the local
pieces of flat space--time. Because theses pieces differ in
non--uniform flows, light propagation in a moving medium
resembles light propagation in curved space--time, {\it i.e.}\ the
medium appears to light as an effective gravitational field
\cite{Gordon}-\cite{Quan2}.

One could conceive of employing moving media for creating
artificial astronomical objects in the laboratory. For example,
water going down the drain of a bathtub appears to light as a
rotating black hole. However, in order to observe spectacular
effects of general relativity in laboratory--based analogues,
truly astronomical flow velocities are required. For establishing
a black hole, the fluid should move faster than the speed of
light in the medium ($c$ divided by the refractive index $n$).
The chances of creating analagoes of black holes are much better
when one employs sound instead of light in appropriate supersonic
flows \cite{U1}-\cite{G2}. One of the most fascinating effects of
black holes is Hawking radiation \cite{Hawking1}-\cite{BD}, the
spontaneous generation of photon pairs near the hole's horizon
where one of the photons falls into the hole and the other
tunnels out of the attraction zone and become visible. The
acoustical equivalent of Hawking radiation is quantum sound in
superfluids \cite{U1}-\cite{G2}. However, superfluidity tends to
break down before the fluid has a chance to move faster than the
speed of sound \cite{LL9}. Furthermore, quantum sound and flow
are just two aspects of the same object, the quantum liquid. The
flow is the macroscopic and the sound the microscopic motion.
Under extreme circumstances such as near the horizon of a sonic
black hole, a clear distinction between sound and flow might be
difficult. An advantage of light in a moving medium is the clear
separation between wave and flow, and both light and medium can
be regarded as separate quantum systems.

Recently, light has been slowed down dramatically
\cite{Hau}-\cite{Budiker} due to an effect called
Electromagnetically Induced Transparency
\cite{Knight}-\cite{Scully}. The use of slow light could open the
opportunity for observing the radiation of quantum--optical black
holes. Note, however, that slow light is a more complicated
phenomenon than light in a moving non--dispersive dielectric
where the refractive index does not depend on the frequency of
light. Slow light is based on a highly dispersive medium created
by dressing the atoms of the medium with an appropriate light
beam. Instead of the phase velocity only the group velocity is
reduced. Therefore, it is not always advisable to conclude
directly from the behavior of light in ordinary dielectrics to
the motional effects of slow--light media. It is the purpose of
these notes to clarify the most elementary features of slow light
in a moving medium \cite{LPliten} and, whenever possible, to give
an instructive simplified picture.

\section{Dispersion relation}

Imagine the medium as being decomposed into drops. Each drop
should be small enough such that the flow does not vary
significantly within the size of the drop, but each drop should
be sufficiently large to sustain several optical oscillations. We
thus assume that the wave length of light is much shorter than
the typical scale of changes in the flow. In this case, we can
describe light by a dispersion relation in the local co--moving
frame of each medium drop denoted with primes
\begin{equation}
\label{dr0}
k'^2 - \frac{\omega'^2}{c^2} -
\chi(\omega')\,\frac{\omega'^2}{c^2} = 0 \,\,.
\end{equation}
The susceptibility $\chi(\omega')$ characterizes the medium. In
Electromagnetically Induced Transparency
\cite{Knight}-\cite{Scully} a coupling beam dresses the upper two
levels of a three--level atom shown schematically in Fig.\ 1.
\begin{figure}
 \begin{center}
  \includegraphics[width=8cm]{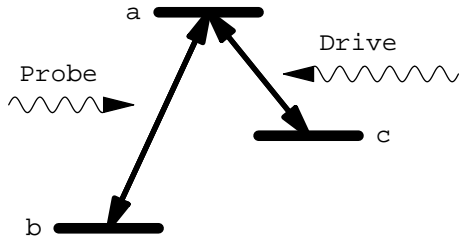}
\caption{Three level system needed for the creation of
electromagnetically
  induced transparency. A strong drive laser field couples the levels
  a and c, making the  medium  transparent for the weak probe laser
  tuned to the transition a$\leftrightarrow$b. }
 \label{levels}
 \end{center}
\end{figure}
The dressing of the upper two levels influences strongly the
propagation of the probe light with a frequency that should match
the atomic transition frequency $\omega_0$ between the lower and
one of the upper levels. Exactly on resonance, the medium
decouples from the probe light and becomes transparent, whereas
ordinary dielectrics are extremely absorbing at atom--light
resonances. Furthermore, the susceptibility changes rapidly, see
Fig.\ 2,
\begin{figure}
 \begin{center}
  \includegraphics[width=8cm]{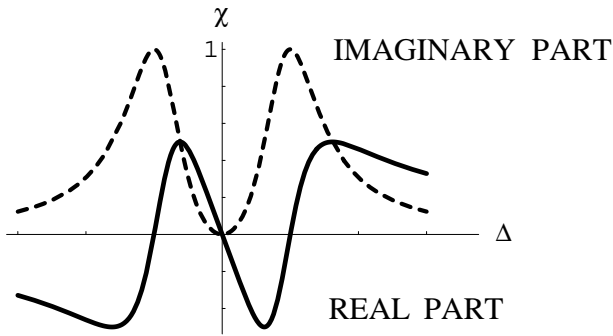}
\caption{Susceptibility for the probe laser beam in
electromagnetically
  induced transparency. The plot shows the dependence of the real
  (whole lines) and imaginary (dashed lines)
parts of the susceptibility $\chi$ on
  the detuning $\Delta$ of the probe beam. The drive beam is assumed
  to be on resonance. Arbitrary units are used. }
 \label{susceptibility}
 \end{center}
\end{figure}
and, near the resonance, $\chi(\omega')$ assumes a linear
dependence on the detuning between the light frequency $\omega'$
and the atomic frequency $\omega_0$,
\begin{equation}
\label{chi} \chi(\omega') = \frac{2c}{\omega_0 v_g
}\,(\omega'-\omega_0) \,\,,
\end{equation}
giving rise to a very low group velocity $v_g$. If the slow--light
medium is moving coherently, the motion will slightly detune each
atom from exact resonance due to the Doppler effect.

\begin{figure}
 \begin{center}
  \includegraphics[width=8cm]{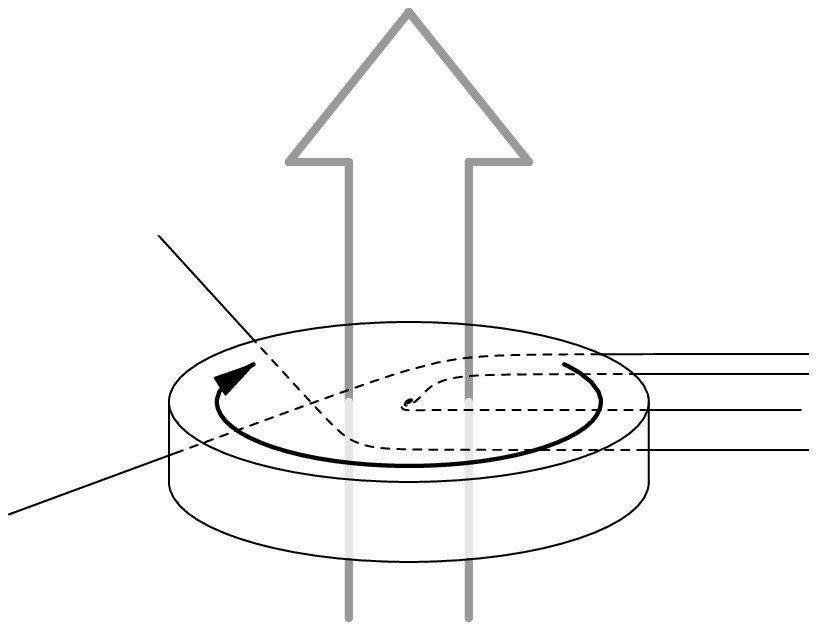}
\caption{Slow light in moving media. The medium should move
orthogonally to the coupling beam (grey arrow). This beam prepares
the medium such that the group velocity of light is significantly
reduced. The probe beam (lines) experiences the Doppler detuning
due to the moving medium and becomes deflected or even trapped.}
 \label{setup}
 \end{center}
\end{figure}

Imagine that the flow of the medium, ${\bf u}$, varies only in
two dimensions and that the coupling beam propagates orthogonally
to the flow, see Fig\ 3. In this case we can ignore the
first--order Doppler effect of the coupling beam and we should
focus only on the Doppler detuning of the probe. In the
laboratory frame the probe shall be monochromatic at the atomic
frequency $\omega_0$. In the co--moving local frames of the
medium the probe--light frequency is Doppler--shifted with the
detuning
\begin{equation}
\omega' - \omega_0 = -{\bf u} \cdot {\bf k} \,\,,
\end{equation}
to first order in $u/c$. Let us transform the dispersion relation
(\ref{dr0}) to the laboratory frame. We take advantage of the
Lorentz invariance of $k^2-(\omega/c)^2$ and obtain to first order
in $u/c$ the relation
\begin{equation}
\label{dr1} k^2 - k_0^2 + 2\,\frac{k_0}{v_g}\,{\bf u} \cdot {\bf
k} = 0
\end{equation}
with
\begin{equation}
k_0 = \frac{\omega_0}{c} \,\,.
\end{equation}
In Ref.\ \cite{LPliten} a more complicated dispersion relation has
been derived which is correct up to second order in $u/c$. Note
that the relation (\ref{dr1}) describes effects in leading order
of $u/(v_g c)^{1/2}$, because the second--order corrections of
Ref.\ \cite{LPliten} are proportional to $v_g^{-1}$.

\section{Magnetic model and metric}

We are going to use the simplified relation (\ref{dr1}) to
illuminate the most characteristic features of slow light in
moving media. First we note that we can write the relation as
\begin{equation}
\label{dr2} \left( {\bf k} + k_0\, \frac{\bf u}{v_g} \right)^2 -
k_0^2\frac{u^2}{v_g^2} = k_0^2 \,\,.
\end{equation}
Monochromatic light waves with fixed polarization thus obey the
wave equation
\begin{equation}
\label{wave} \left( -i\nabla + k_0\, \frac{\bf u}{v_g} \right)^2
\phi - k_0^2\frac{u^2}{v_g^2}\,\phi = k_0^2\,\phi \,\,.
\end{equation}
The flow has a two--fold effect: On one hand, the velocity ${\bf
u}$ appears as an effective vector potential, for example the
magnetic vector potential acting on an electron wave \cite{LL3},
and, on the other hand, the hydrodynamic pressure proportional to
$u^2$ \cite{LL6} acts as a scalar potential. This resembles the
R\"ontgen effect of static electromagnetic fields on polarizable
atoms \cite{Wei,LW}.

Alternatively, we can regard the flow as generating an effective
gravitational field. To see this, we introduce the
four--dimensional wave vector
\begin{equation}
k_\nu = (k_0, -{\bf k}) \,\,,
\end{equation}
and, adopting Einstein's summation convention, write the
dispersion relation (\ref{dr1}) as
\begin{equation}
g^{\mu\nu} k_\mu k_\nu = 0
\end{equation}
with the contravariant metric tensor
\begin{equation}
g^{\mu\nu} = \left(
\begin{array}{cc}
1 & {\displaystyle {\bf u}/{v_g}} \\
{\displaystyle {\bf u}/{v_g}} & -{\bf 1}
\end{array}
\right) \,\,.
\end{equation}
Four--dimensional wave vectors are null vectors with respect to
the effective metric $g^{\mu\nu}$ and, in turn \cite{LPstor},
light rays follow zero-geodesics with respect to the line element
\begin{equation}
ds^2 = g_{\mu\nu} dx^\mu dx^\nu \quad,\quad dx^\nu = (ct, {\bf
x}) \,\,.
\end{equation}
The covariant metric tensor $g_{\mu\nu}$ is the inverse of the
contravariant one, and is given by
\begin{equation}
g_{\mu\nu} = \frac{1}{1 + {u^2}/{v_g^2}} \left(
\begin{array}{cc}
1 & {\displaystyle {\bf u}/{v_g}} \\
{\displaystyle {\bf u}/{v_g}} & -{\bf 1}
\end{array}
\right) \,\,.
\end{equation}
In contrast to sound \cite{Visser} or light in non--dispersive
media \cite{LPstor}, a genuine event horizon of slow light cannot
exist at this level of approximation. The potential existence of
an event horizon is thus confined to effects to second order in
$u/(v_g c)^{1/2}$which are excluded from our simplified model
\cite{Vissercomment}.

\section{Non--relativistic analogue}

To get more insight into the behavior of slow light in moving
media, let us study another analogue. Consider a
non--relativistic free particle attached to moving frames. The
particle is supposed to move freely in each of the local frames
but is bound to adapt from one frame to the next, similar to
light in drops of flowing water. However we use pure
non--relativistic physics to transform from frame to frame. In
case the local frames constitute a global frame, for example a
rotating solid body, the particle would move along a straight
line with respect to this frame. In the laboratory frame, the
particle's trajectory appears to be bent, attributed to the
effect of Coriolis and centrifugal forces. The Lagrange function
of such a fictitious non--relativistic particle is
\begin{equation}
L = \frac{1}{2}\,{v'}^2 = \frac{1}{2}\,({\bf v} - {\bf w})^2 \,\,,
\end{equation}
up to a constant mass factor that we can set to unity, because we
are interested in inertial effects only. The velocity of the
local intertial frames is ${\bf w}$ and we have used the
non--relativistic addition theorem of velocities
\begin{equation}
{\bf v}' = {\bf v} - {\bf w} \,\,.
\end{equation}
Note that we could apply Einstein's relativistic addition theorem
of velocities to describe light in moving non--dispersive media of
refractive index $n$, but we should use an effective speed of
light of $c/n$ in the theorem. The momentum of the fictitious
non--relativistic particle is
\begin{equation}
\label{p1} {\bf p} \equiv \frac{\partial L}{\partial {\bf v}} =
{\bf v} - {\bf w}
\end{equation}
and the Hamiltonian is
\begin{equation}
\label{h} H \equiv {\bf v}\cdot{\bf p} - L = \frac{1}{2}\,p^2 -
{\bf w}\cdot{\bf p} = E \,\,.
\end{equation}
The important point is that we can translate the Hamiltonian
(\ref{h}) into the dispersion relation (\ref{dr1}) by setting
\begin{equation}
\label{setting} {\bf p} = c\,\frac{\bf k}{k_0} \,\,,\quad {\bf w}
= \frac{c}{v_g}\,{\bf u} \,\,,\quad E = \frac{1}{2}\,c^2 \,\,.
\end{equation}
with the enhanced effective frame velocity ${\bf w}$
\cite{LPgyro}. Consequently, slow light experiences the moving
medium in the same way as a non--relativistic particle
experiences moving inertial frames. Excused by its fictitious
nature, the ``non--relativistic'' particle would move at an
extraordinary speed reaching $c$ in regions where the medium is
at rest, because the energy is
\begin{equation}
\frac{1}{2}\,c^2 = E = H = \frac{1}{2}\,v^2 - \frac{1}{2}\,w^2
\,\,.
\end{equation}
We obtain from Hamilton's equations,
\begin{equation}
{\bf v} = \dot{\bf r} = \frac{\partial H}{\partial {\bf p}}
\,\,,\quad \dot{\bf p}= -\frac{\partial H}{\partial {\bf r}} \,\,,
\end{equation}
the equation of motion of the fictitious particle
\begin{equation}
\dot{\bf v} = (\nabla\times{\bf w})\times{\bf v} +
\frac{1}{2}\,\nabla w^2 \,\,.
\end{equation}
The first term describes the Coriolis and the second the
centrifugal force.

\section{Light rays around a vortex}

Consider a specific example of a non--uniform flow, a vortex.
Quantum vortices in alkali Bose--Einstein condensates have been
recently made \cite{Matthews,Madison}. They give rise to
intriguing slow-light phenomena. However, vortices do not posses
genuine event horizons \cite{Vissercomment} and they will not
radiate spontaneously. The flow of an ideal vortex of vorticity
${\cal W}$ is
\begin{equation}
\label{vortex} {\bf u} = \frac{\cal W}{r}\,{\bf e}_\varphi =
i\,\frac{\cal W}{z^*}\,\,,
\end{equation}
written in polar coordinates or using the complex notation
$z=x+iy$, $z^*=x-iy$ of the planar Cartesian coordinates $x,y$.
The curl of the vortex flow vanishes,
\begin{equation}
\nabla \times {\bf u} = {\bf 0} \,\,,
\end{equation}
and hence we obtain the equation of motion of a particle in an
attractive $r^{-2}$ potential \cite{LL1}-\cite{Denschlag},
\begin{equation}
\dot{\bf v} = \frac{1}{2}\,\nabla w^2 = -\,\frac{c^2{\cal
W}^2}{v_g^2}\,\frac{\bf r}{r^4}\,\,,
\end{equation}
or, in complex notation,
\begin{equation}
\label{z} \ddot{z} + \frac{c^2{\cal W}^2}{v_g^2}\,\frac{z
}{|z|^4} = 0 \,\,.
\end{equation}
The trajectories of the fall into a $r^{-2}$ singularity are well
known in polar coordinates \cite{LL1} but here we suggest a more
elegant description using the complex notation of the planar
Cartesian coordinates,
\begin{equation}
z = \left(-ct + i\,\frac{b}{\mu}\right)^{(1+\mu)/2} \left(-ct -
i\,\frac{b}{\mu}\right)^{(1-\mu)/2}
\end{equation}
with the real parameter $b$ and
\begin{equation}
\mu = \left(1-\frac{{\cal W}^2}{v_g^2 b^2}\right)^{-1/2} \,\,.
\end{equation}
One can easily verify that this solution satisfies the equation
of motion (\ref{z}). In the infinite past, $t\rightarrow
-\infty$, the trajectories approach
\begin{equation}
z \sim -c t + i b \,\,.
\end{equation}
From this asymptotics we see that the light is incident from the
right to the left with initial velocity $c$ and impact parameter
$b$. Depending on the impact parameter, the coefficient $\mu$ is
real or purely imaginary. In the case of a real $\mu$ when ${\cal
W}^2 < v_g^2 b^2$ the incident light particle will be able to
escape, because the modulus squared of $z$, $c^2 t^2 +
b^2/\mu^2$, will never reach zero, the vortex core. In the
infinite future, $t\rightarrow +\infty$, the particle will
approach
\begin{equation}
z \sim c t \, (-1)^\mu + i b = c t e^{i\pi\mu} + i b
\end{equation}
and, consequently, leaves at the angle $\pi\mu$. In the case of a
purely imaginary coefficient $\mu$ when ${\cal W}^2 > v_g^2 b^2$
the light is sucked into the vortex, because the modulus squared
of $z$, $c^2 t^2 + b^2/\mu^2$, reaches zero at the time
$t=-|b/(\mu c)|$. In practice, the light will hit the vortex core
and will bounce back. In theory, the fictitious light particle
disappears from this world and reappears on another Riemann
sheet. In the borderline case when ${\cal W}^2 = v_g^2 b^2$ the
parameter $\mu$ tends to infinity and the trajectory approaches
\begin{equation}
\label{limitspiral} z = -c t \lim_{\mu\rightarrow\infty} \left(1 -
\frac{1}{\mu}\,\frac{i b}{c t}\right)^\mu = -c t \,
\exp\left(-\frac{i b}{c t}\right) = \exp\left(\pm\frac{i {\cal
W}}{v_g c t}\right)\,\,.
\end{equation}
Light falling into the vortex core describes distinctive spirals
illustrated in Fig.\ 4.
\begin{figure}
 \begin{center}
  \includegraphics[width=8cm]{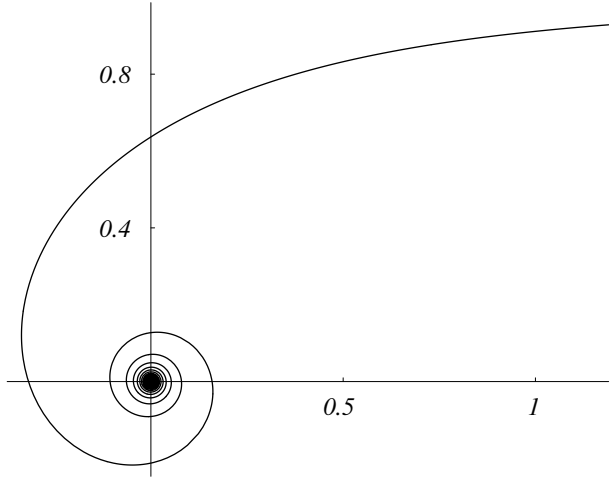}
\caption{Light rays spiraling towards the vortex core. The plot
shows the limiting spiral of described in Eq.\
(\ref{limitspiral}).}
 \label{spiral}
 \end{center}
\end{figure}

\section*{Light waves around a vortex}

Near the vortex core, rays of slow light feel the presence of the
centrifugal $r^{-2}$ attraction but no Coriolis force, because
the curl of the flow vanishes. Light waves, however, will show a
distinctive interference pattern, an optical Aharonov--Bohm
effect \cite{Hannay}-\cite{ABbook} due to the long--range nature
of a vortex flow. Far outside the vortex core, light rays are
hardly deflected, yet rays passing the vortex in flow direction
will be dragged and those swimming against the current lag
behind. This does not affect the ray trajectories but it produces
a phase shift between the dragged and the lagging rays, and, in
turn, creates a typical interference pattern \cite{AB,ABbook}.
Note that the Aharonov--Bohm effect of waves in moving media is
not restricted to light. Indeed, Berry {\it et al.}
\cite{Berryetal} have reported and analyzed a beautiful bathtub
experiment with water waves in a tank, visualizing clearly the
interference structure generated by the Aharonov-Bohm effect. A
further experiment has been reported \cite{Vivanco} showing spiral
waves. Acoustical analogues of the effect have been observed in
moving classical media \cite{Roax} and are predicted for
superfluids \cite{Davidowitz}. The acoustical effect might even
lead to a friction felt by traveling vortices due to the so-called
Iordanskii force \cite{Iordanskii}-\cite{Stone}. The optical
Aharonov--Bohm effect of slow light \cite{LPliten} can be applied
to observe {\it in situ} the flow of quantum vortices in alkali
Bose--Einstein condensates \cite{Matthews,Madison} using
phase--contrast imaging \cite{Andrews}.

Consider the propagation of slow--light waves through a vortex
flow. In polar coordinates the wave equation (\ref{wave})  is
\begin{equation}
\label{vortexwave} \left[ \frac{\partial^2}{\partial r^2} +
\frac{1}{r} \frac{\partial}{\partial r} + \frac{1}{r^2} \left(
\frac{\partial}{\partial \varphi} + i \nu_{AB} \right)^2 +
\frac{\nu_{AB}^2}{r^2}\right] \phi = k_0^2\, \phi
\end{equation}
with the optical Aharonov--Bohm flux quantum
\begin{equation}
\nu_{AB} = k_0\,\frac{\cal W}{v_g} \,\,.
\end{equation}
We could simply eliminate the optical analogue of the vector
potential by representing $\phi$ as
\begin{equation}
\label{ansatz} \phi = \phi_0\,\exp(-i\nu_{AB}\varphi)
\end{equation}
where $\phi_0$ feels only the local $r^{-2}$ attraction due to
the centrifugal force. The modulation $\exp(-i\nu_{AB}\varphi)$
describes the long--range Aharonov--Bohm phase pattern \cite{AB}.
Strictly speaking, however, the ansatz (\ref{ansatz}) is only
justified when the function $\exp(-i\nu_{AB}\varphi)$ is
single--valued after a complete cycle of $\varphi$, {\it i.e.}
when $\nu_{AB}$ is an integer. For non--integer flux quanta the
interference pattern is slightly more complicated, showing, most
prominently, a line of zeros of the wave function after passing
the vortex (which resolves the problem of a multi--valued phase,
because a phase is not defined when the amplitude is zero). Even
in the case of non--integer optical flux quanta $\nu_{AB}$, the
modulation $\exp(-i\nu_{AB}\varphi)$ accounts for the dominant
phase pattern.

The correct positive--frequency component of a slow--light wave
incident from the right is an appropriate superposition \cite{LW}
\begin{equation}
\phi = \sum_{m=-\infty}^{+\infty} \phi_m
\end{equation}
of the partial waves
\begin{equation}
\label{psim} \phi_m = u_m(r)\,\exp(i m \varphi - i\omega_0 t)
\,\,,\quad u_m = (-i)^\nu \,J_{\nu}(k_0 r)
\end{equation}
characterized by the Bessel functions $J_\nu$ with the index
\begin{equation}
\label{index} \nu = \sqrt{(m+\nu_{AB})^2-\nu_{AB}^2} \,\,.
\end{equation}
Figure 5 shows plots of slow--light waves around a vortex to
illustrate the long--range phase shift brought about by the
optical Aharonov--Bohm effect.
\begin{figure}[htbp]
  \begin{center}
    \includegraphics[width=20.5pc]{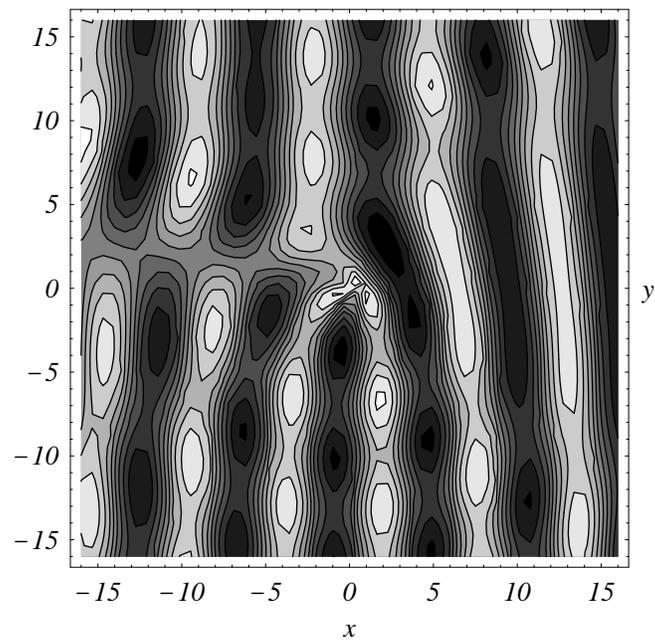}
    \vspace*{2mm}

    \caption{Slow-light waves traveling through a vortex.
    The dragging effect of the vortex shifts the phase of the
    incident light, depending on whether the light propagates
    with or against the flow.}
    \label{fig:vortex}
  \end{center}
\end{figure}

Close to the vortex core, a light ray is doomed to fall into the
singularity when the square of the impact parameters, $b^2$, does
not exceed ${\cal W}^2/v_g^2$. In this case the angular momentum
$m$ lies in the interval
\begin{equation}
\label{trap} -2\nu_{AB} \le m \le 0 \,\,,
\end{equation}
because, according to Eqs.\ (\ref{p1},\ref{setting}) and the
pattern (\ref{vortex}) of the vortex flow,
\begin{equation}
{\bf r}\times{\bf k} = {\bf r}\times k_0({\bf v}-{\bf w})/c =
(k_0 b - \nu_{AB})\,{\bf e}_z \,\,.
\end{equation}
The fall of light rays into the vortex core corresponds to a
drastic change of the behavior of the corresponding partial waves
$\phi_m$. In the case (\ref{trap}) the index (\ref{index}) of the
Bessel function in a radial wave (\ref{psim}) is purely
imaginary. Using the first term in the power--series expansion of
the Bessel functions \cite{Erdelyi} we see that $u_m$ is rapidly
oscillating near the core,
\begin{equation}
u_m \sim \frac{(-i)^\nu}{\Gamma(\nu+1)}\,\left(\frac{k_0
r}{2}\right)^\nu = \frac{(-i)^\nu}{\Gamma(\nu+1)}\,\exp\left[i{\rm
Im}\,\nu\,\ln\left(\frac{k_0 r}{2}\right)\right] \,\,.
\end{equation}
This describes a rapid flow of light towards the centre of
attraction, if ${\rm Im}\,\nu$ is negative and an outgoing wave
for the positive branch. Far outside the vortex core,
$r\rightarrow\infty$, the radial waves approach
\begin{equation}
\label{asymp} u_m \sim \frac{1}{\sqrt{2\pi k_0
r}}\,\left[\exp\left(i k_0 r - i\pi\nu - i\frac{\pi}{4}\right) +
\exp\left(-i k_0 r + i\frac{\pi}{4}\right)\right]\,\,,
\end{equation}
as we see from the asymptotics of the Bessel functions
\cite{Erdelyi}. For a negative imaginary $\nu$ the second term in
Eq.\ (\ref{asymp}) dominates and so most of the light is incident,
yet a small fraction with weight $\exp(-2i\pi\nu)$ is able to
tunnel out of the attraction zone.

Where is the borderline between pure influx and partial tunneling
out of the attraction zone? Consider the geometrical optics of a
radial wave. We make the eikonal ansatz
\begin{equation}
\label{super} u_m = c_m^+u_m^+ + c_m^-u_m^- \,\,\quad u_m^\pm =
|u_m|\,\exp(\pm i R)\,\,,
\end{equation}
regard $|u_m|$ as a slowly varying envelope, and obtain from the
wave equation (\ref{vortexwave}) the radial eikonal
\begin{equation}
\label{eikonal} R = i\nu\left[\sqrt{\rho^2+1} - {\rm
arsinh}\left(\frac{1}{\rho}\right) + i\frac{\pi}{2}\right]
\,\,,\quad \rho = \frac{k_0 r}{i\nu} \,\,.
\end{equation}
Quite typically, valuable insight into the behavior of waves can
be gained by analytical continuation to the complex plane of the
wave's argument. Consider the line in the complex plane of the
radius $r$ where the eikonal is purely imaginary, {\it i.e.} where
\begin{equation}
{\rm Re}\,\left[\sqrt{\rho^2+1} - {\rm
arsinh}\left(\frac{1}{\rho}\right) \right] = 0 \,\,,
\end{equation}
see Fig.\ 6. This line is called a Stokes line
\cite{Furry,Ablowitz}. When passing a Stokes line, one component
of the superposition (\ref{super}) becomes exponentially small
and hence irrelevant \cite{Furry}. The asymptotics in the limit of
geometrical optics switches from a two--component regime, for
example in--and--out flux, to a single component behavior, such
as pure influx \cite{Furry}. The Stokes line in Fig.\ 6 thus marks
the boundary where light is able to tunnel out of the optical
black hole.
\begin{figure}[htbp]
  \begin{center}
    \includegraphics[width=20.5pc]{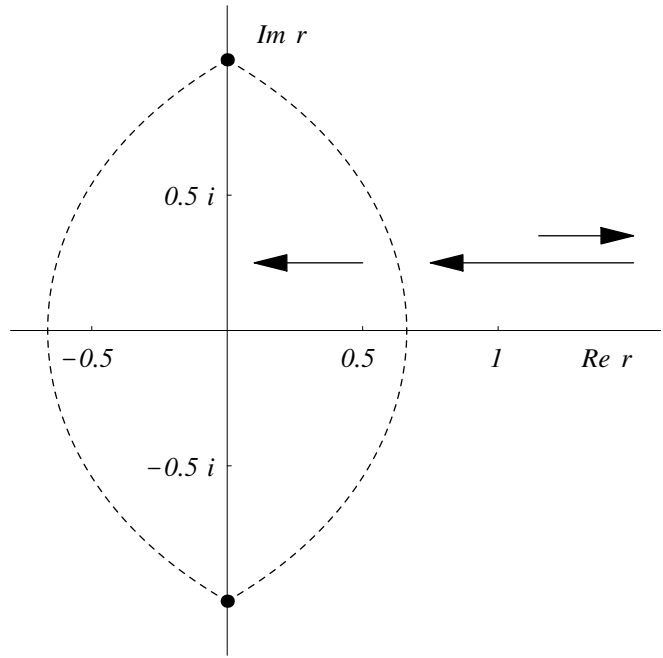}
    \vspace*{2mm}

    \caption{Stokes lines of an optical black hole. Stokes lines
    \cite{Furry}
    are lines in the complex plane of the radius $r$
    where the radial eikonal of light waves is purely imaginary.
    The lines connect the turning points of light rays which are
    complex in the general case of trapping (and purely imaginary
    for an optical black hole). The Stokes lines are calculated
    from the radial part of the eikonal (41) and are displayed in
    the scaled units employed there. The arrows indicate the light
    flux in the regions bounded by a Stokes line.}
    \label{fig:stokesline}
  \end{center}
\end{figure}

\section{Summary}
We analyzed the propagation of slow light in moving media in the
case when the light is monochromatic in the laboratory frame.
Slow light is generated by the electromagnetically induced
transparency of an atomic transition. The Doppler detuning due to
the moving medium generates flow-dependent effects. We considered
analogies to magnetism and general relativity and studied light
propagation around a vortex in some detail. The subject is far
from being exhausted --- light detuned from the atomic resonance
and non--monochromatic light will behave differently, opening
perhaps new exciting analogies and opportunities for laboratory
tests of astronomical effects on Earth.

\section{Acknowledgements}

U.L. acknowledges the support of the Alexander von Humboldt
Foundation and of the G\"oran Gustafsson Stiftelse during his
stay at the Royal Institute of Technology. P.P. was partially
supported by the research consortium {\it Quantum Gases} of the
Deutsche Forschungsgemeinschaft.

\end{document}